\documentclass[twocolumn,nolinenumbers]{aastex631}
\usepackage{graphicx}
\usepackage{dcolumn}
\usepackage{bm}
\usepackage{amssymb}
\usepackage{amsmath}

\usepackage{epstopdf}

\begin{document}
\title{High-energy Neutrinos from Outflows Powered by Kicked Remnants of Binary Black Hole Mergers in AGN Accretion Disks}

\author[0009-0007-0717-3667]{Zhi-Peng Ma}
\affiliation{Department of Astronomy, School of Physics, Huazhong University of Science and Technology, Wuhan, Hubei 430074, China}

\author[0000-0003-4976-4098]{Kai Wang}
\affiliation{Department of Astronomy, School of Physics, Huazhong University of Science and Technology, Wuhan, Hubei 430074, China}

\correspondingauthor{Kai Wang}
\email{kaiwang@hust.edu.cn}


\begin{abstract}
 Merging of stellar-mass binary black holes (BBH) could take place within the accretion disk of active galactic nuclei (AGN). The resulting BH remnant is likely to accrete the disk gas at a super-Eddington rate, launching a fast, quasi-spherical outflow (wind). Particles will be accelerated by shocks driven by the wind, subsequently interacting with the shocked disk gas or radiation field through hadronic processes and resulting in the production of high-energy neutrinos and potential electromagnetic (EM) emissions. This study delves into the intricate evolution of the shock driven by the remnant BH wind within AGN disks. Subsequently, we calculated the production of neutrinos and the expected detection numbers for a single event, along with their contributions to the overall diffuse neutrino background. Our analysis, considering various scenarios, reveals considerable neutrino production and possible detection by IceCube for nearby events. The contribution of the remnant BH winds on the diffuse neutrino background is minor due to the low event rate density, but it can be improved to some extent for some optimistic parameters. We also propose that there could be two neutrino/EM bursts, one originating from the premerger BBH wind and the other from the remnant BH wind, with the latter typically having a time gap to the GW event of around tens of days. When combined with the anticipated gravitational waves (GW) emitted during the BBH merger, such a system emerges as a promising candidate for joint observations involving neutrinos, GWs, and EM signals.

\end{abstract}

\keywords{Neutrino astronomy --- High energy astrophysics --- Active galactic nuclei --- Gravitational waves --- Black holes}

\section{Introduction} \label{sec:intro}
It is believed that many stars and stellar remnants can be embedded within the accretion disk of AGN~\citep{artymowicz1993star,cantiello2021stellar,dittmann2021accretion,collin1999star,fabj2020aligning,wang2011star}. The abundant stellar remnants would migrate inward the disk and be trapped at the specific orbit~\citep{bellovary2016migration,tagawa2020formation}, leading to the high possibility of compact object mergers~\citep{cheng1999formation,mckernan2020black}, such as binary neutron star (BNS) mergers~\citep{mckernan2020black}, binary black hole (BBH) mergers, and NS-BH (NSBH) mergers~\citep{bartos2017rapid,leigh2018rate,kimura2021outflow,tagawa2022can}. Especially, among these compact object mergers in the AGN disk, BBH mergers have attracted the most attention since they have been prime targets for Earth-based gravitational wave (GW) detectors. Another important point is that BBH mergers embedded in AGN disks can potentially power their electromagnetic counterparts by accreting the abundant disk gas~\citep{wang2021accretion,kimura2021outflow,tagawa2022can,2024ApJ...961..206C}, while for the normal environment, the accompanying EM radiations with BBH mergers are not expected because there are no accretion materials to power a jet. Recently, seven AGN flares have been suggested to have a statistical association with LIGO/Virgo BBH merging events~\citep{graham2023light}. In addition, the Zwicky Transient Facility (ZTF) has reported an optical counterpart candidate to the GW event GW190521~\citep{graham2020candidate}, which comes from the most massive stellar-mass BBH system to date (up to $150M_{\odot}$) ~\citep{abbott2020gw190521}.

Besides the GW and EM signals, astrophysical activities within AGN disks, e.g., supernovae~\citep{zhu2021thermonuclear,Zhou2023a}, gamma-ray bursts~\citep{zhu2021high}, stellar-mass BH jets~\citep{Tagawa_2023} and the remnant BH jets of BBH mergers~\citep{zhou2023b}, can also be the potential high-energy neutrino source. For the BBH system embedded in AGN disks, the merged BH remnant will be kicked into an intact AGN disk environment. Then, the surrounding disk gas will be accreted with a super-Eddington accretion rate, and the jet will be launched by Blandford–Znajek (BZ) mechanism. The interaction between the postmerger BZ jet and the AGN disk materials will create diverse shocks, accelerate particles, and produce neutrinos by hadronic processes~\citep{zhou2023b}. In addition to the jet, the remnant BH of the BBH post-merger will also power the radiation-driven wind~\citep{jiang2014global,ohsuga2005supercritical,skadowski2013energy}, and this wind interacting with the disk medium may also play an important role in multimessenger signals of BBH mergers in AGN disks.


\begin{figure*}
\label{fig_phy}
\begin{center}
\includegraphics[width=0.8\linewidth]{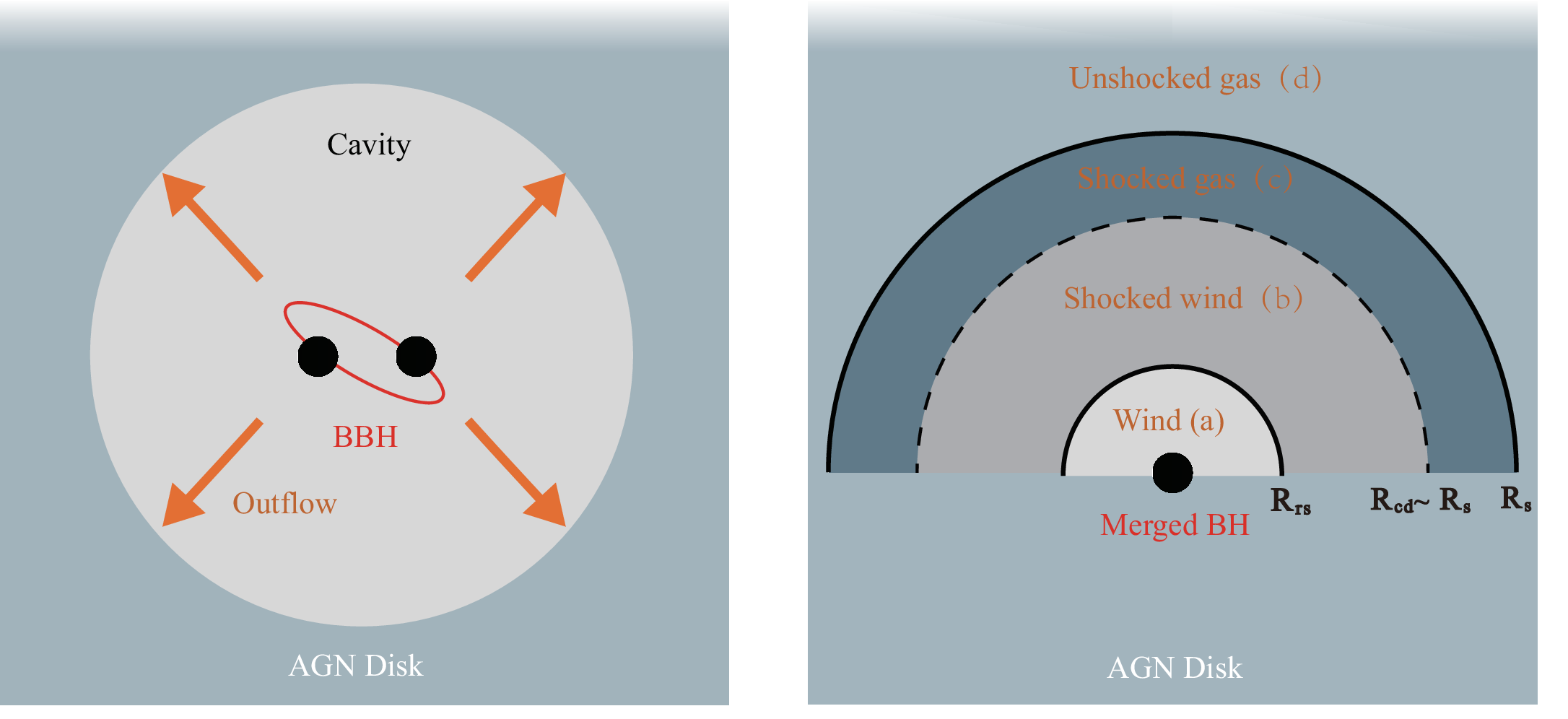}
\caption{Schematic picture for the system. Left: The BBH wind will flow out the ambient gas and create a cavity around the merged BH. Right: When the merged BH is kicked out of the cavity into a new disk environment, it will accrete the disk gas and power a new wind, and this wind will produce a forward shock sweeping across the disk gas and a reverse shock sweeping across the wind, dividing the system into four regions. See texts for more details.}
\label{fig_dyn}
\end{center}
\end{figure*}


BBH mergers are the potential GW-neutrino association source, and \cite{zhou2023b} estimate the joint detection of GW from BBH merger and high-energy neutrinos from the jet of the postmerger remnant BH within AGN disks. Compared with the jet of remnant BH, the sub-relativistic wind driven by the merged BH is quasi-isotropic without the beaming effect, leading to a more promising detection of GW-neutrino association for the current small number of detected BBH mergers. In this paper, we study the neutrino emission from the interaction between the wind driven by the remnant BH and the disk medium. In our scenario, the premerger BBH system will produce a radiation-driven wind, and create a cavity around the merged BH in the AGN disk. When the merged BH is kicked into a new, gas-abundant disk environment, it will continue to accrete the disk gas and power a new material wind~\citep{kimura2021outflow}. Such a wind will generate a forward shock in the disk gas, hence particles can be accelerated at the shock front. Then, the accelerated protons will interact with the disk materials and thermal radiations of the shocked gas to produce high-energy neutrinos. We show a schematic picture of our scenario in Fig.~\ref{fig_phy}.

This paper is organized as follows. We show the equations and solutions for shock dynamics in Section~\ref{sec_dyn}. The method for estimating neutrino emission is shown in Section~\ref{sec_np}. The neutrino energy fluence, expected numbers, and contributions to diffuse neutrino background are shown in Section~\ref{sec_res}. We summarize our work in Section~\ref{sec_sum}.

\section{Dynamical model} 
\label{sec_dyn}
In this section, we describe the properties of the disk, including density profile and temperature distribution. Then, we set up the dynamic equations for the forward shock and illustrate the solutions. Notice that we only focus on the outflow driven by the merged BH. However, such equations for shock dynamics and neutrino productions are also valid for the premerger outflow and the massive BH formed through other ways (such as Core-collapsed).

\subsection{Disk Model and Outflow Luminosity}
Since BBH and the final merged BH would migrate within the disk of the supermassive black hole (SMBH), we set the distance $r$ from the merged BH to SMBH as a free parameter, in the range of $10^3R_{\rm sch}\le r \le 10^4R_{\rm sch}$, where $R_{\rm sch}=2GM_{\rm SMBH}/c^2$ is the  Schwarzschild radius of SMBH, G is the
gravitational constant, $M_{\rm SMBH}$ is the SMBH mass, and c is the speed of light. We adopt a modified $\alpha$-disk model~\citep{cantiello2021stellar} to calculate the gas temperature $T_{\rm mid}$ and gas density profile $\rho_{\rm mid}$ for the disk midplane, also the scale height H and surface density $\Sigma$ for the disk. By considering gravitational stability, all these physical quantities have different expressions in the inner and outer regions of the disk. The detailed formulas are given in the Appendix~\ref{sec_apx1}. Once the expressions for the midplane are confirmed, one can get the vertical density profile in Gaussian form~\citep{zhu2021high} (with fixed SMBH mass),
\begin{equation}
    \rho_{\rm disk}(r,h)=\rho_{\rm mid}(r){\rm exp}(-h^2/2H^2),
\end{equation}
where $h=r/H$ is the aspect ratio. For the gas temperature distribution in the vertical direction, we simply assume that $T_{\rm disk}(r,h)=T_{\rm mid}(r)$. The pressure can be calculated by $P_{\rm disk}(r,h)=\rho_{\rm disk}/m_pk_bT_{\rm disk}$ under the assumption of the ideal gas, where $k_b$ is the Boltzmann constant. We note that the pressure of the disk gas may not follow the ideal gas relation, but such an estimation will not make a significant deviation in calculations, because the deceleration of the shocked gas is mainly caused by the gravity of the merged black hole rather than the pressure of the disk gas.

When the merged BH is kicked out of the cavity into a gas-abundant region on the disk, it will accrete the disk gas in the way of Bondi-Hoyle-Littleton (BHL) accretion and power a fast wind. The accretion rate can be estimated through~\citep{bondi1952spherically,hoyle1939effect,edgar2004review}
\begin{equation}
 \dot{M}_{\rm bh}(r)=\frac{4\pi G^2M^2_{\rm bh}\rho_{\rm mid}}{v^3_{\rm kick}},
\end{equation}
where $M_{\rm bh}$ is the mass of the merged black hole, which is fixed as $150M_{\odot}$. $v_{\rm kick}\simeq 1\times 10^7 \rm{cm/s}$ is the kick velocity~\citep{campanelli2007maximum,herrmann2007unequal,gonzalez2007maximum}. The kinetic power of the wind can be estimated through
\begin{equation}
    L_{\rm w}=\eta_{w}\dot{M}_{\rm bh}v^2_{w},
\end{equation}
where $\eta_{w}$ is the energy distribution factor for the BHL accretion rate and $v_{w}$ is the velocity of the wind.

\subsection{Dynamic Equations}
We adopt a 1D shock model whose physical picture is similar to that of stellar wind or AGN wind. We assume a steady spherical wind launched from BH with a velocity $v_{\rm w}=10^{10} \rm cm/s$. The disk gas around BH is also assumed to be spherically symmetric, based on the direction vertical to the midplane. When the fast wind is injected into the disk, a forward shock and a reverse shock will be formed at the radii of $R_{\rm s}$ and $R_{\rm rs}$ respectively. Two shock radii together with the contact discontinuity at radius $R_{\rm cd}$ divide the system into four regions: (a) the unshocked wind moving with $v_{w}$, (b) the shocked winds, (c) the shocked disk gas, and (d) the unshocked disk gas. The schematic picture can be seen in Fig.~\ref{fig_phy}. 

The forward shock will sweep the disk gas in Region (d) and compress it into a shell. For simplicity, we assume that $R_{s}\simeq R_{cd}$ (thin shell approximation). Under such an assumption~\citep{weaver1977interstellar,faucher2012physics,wang2015probing,liu2018can}, the forward shock and the gases in Region (c) have a unified velocity, notated $v_{\rm s}$. For the shocked wind in Region (b), we follow the treatment proposed by~\cite{liu2018can}, which considers a steady shocked wind with homogeneous number density $n_{\rm sw}$, temperature $T_{\rm sw}$ and gas pressure $P_{\rm sw}$ in the region at any time\footnote{Such an assumption is validated when the local sound cross time $\sqrt{P_{\rm sw}/\rho_{\rm sw}}$ is shorter than the system dynamical timescale, which is generally satisfied in our scenario.}. The number density is $n_{\rm sw}=4n_{w}=\dot{M}_{w}/\pi R^2_{\rm rs}m_{\rm p}v_{w}$, where the wind injection rate is $\dot{M}_{w}=2L_{w}/v^2_{w}$. Under the condition of mass conservation, $R^2v_{\rm sw}(R)$ is a constant in Region (b), where $R$ is the distance to the merged BH and $v_{\rm sw}(R)$ is the velocity of shocked wind at radius $R$. There are two boundary conditions for $v_{\rm sw}(R)$ at $R_{\rm s}$ and at $R_{\rm rs}$, which are $v_{\rm sw}(R_{\rm s})=v_{\rm s}$ and $v_{\rm sw}(R_{\rm rs})=v'_{\rm sw}$, where $v'_{\rm sw}$ is the velocity of the shocked wind just behind $R_{\rm rs}$. By combining two boundary conditions with mass conservation, one can find the relation,
\begin{equation}
\label{eq_4}
    v'_{\rm sw}=(\frac{R_{\rm s}}{R_{\rm rs}})^2v_s.
\end{equation}
Together with the Rankine–Hugoniot jump relation $ v_{\rm w}-v_{\rm rs}=4(v'_{\rm sw}-v_{\rm rs})$, where $v_{\rm rs}$ is the velocity of the reverse shock, one can get the useful expression for $v_{\rm rs}$, that is,
\begin{equation}
\label{eq_5}
    v_{\rm rs}=\frac{4}{3}(\frac{R_{\rm s}}{R_{\rm rs}})^2v_{\rm s}-\frac{1}{3}v_{\rm w}.
\end{equation}
Once the evolution of the forward shock is known, one can estimate the motion of reverse shock through Eq.(\ref{eq_5}) directly.

Next, we need to describe the evolution of the forward shock. For the shocked disk gas in Region (c), we have the motion equation,
    \begin{equation}
    \label{eq_mov}
    \frac{d(M_{\rm s}v_{\rm s})}{dt}=4\pi R^2_{\rm s}(P_{\rm tot}-P_{\rm disk})-\frac{GM_{\rm tot}M_{\rm s}}{R^2_{\rm s}},
\end{equation}
where $P_{\rm tot}=P_{\rm sw}+P_{\rm ph}$ is the pressure provided by the gas and photons in Region (b), which pushes the shell forward. The reason for considering the photon pressure is that the radiation from electrons and protons in Region (b) hardly escapes until the shock breaks out, and most of the photons move with the shocked wind and provide pressure together with the matter. $P_{\rm disk}$ is the pressure of the disk gas, which resists expansion of the shell. The second term on the right-hand side represents the gravity exerted by the total gravitational mass within $R_{\rm s}$, which mainly causes the deceleration of the shell, where $M_{\rm tot}=M_{\rm bh}+M_{\rm s}/2$. $M_{\rm s}$ is the total mass of the shocked disk gas, which can be obtained by integrating the mass swept by the forward shock: $M_{\rm s}(R_{\rm s})=\int_{0}^{R_{\rm s}}{4\pi R^2\rho_{\rm disk}(r, R) dR}$. We ignore the effect of gravity of the SMBH, since the Hill radius $R_{\rm Hill}=(M_{\rm bh}/3M_{\rm SMBH})^{1/3}r$ is comparable to the scale height $H$, suggesting that gravity of the merged stellar-mass BH is dominated.
\begin{figure*}
\includegraphics[width=0.5\linewidth,height=0.4\linewidth]{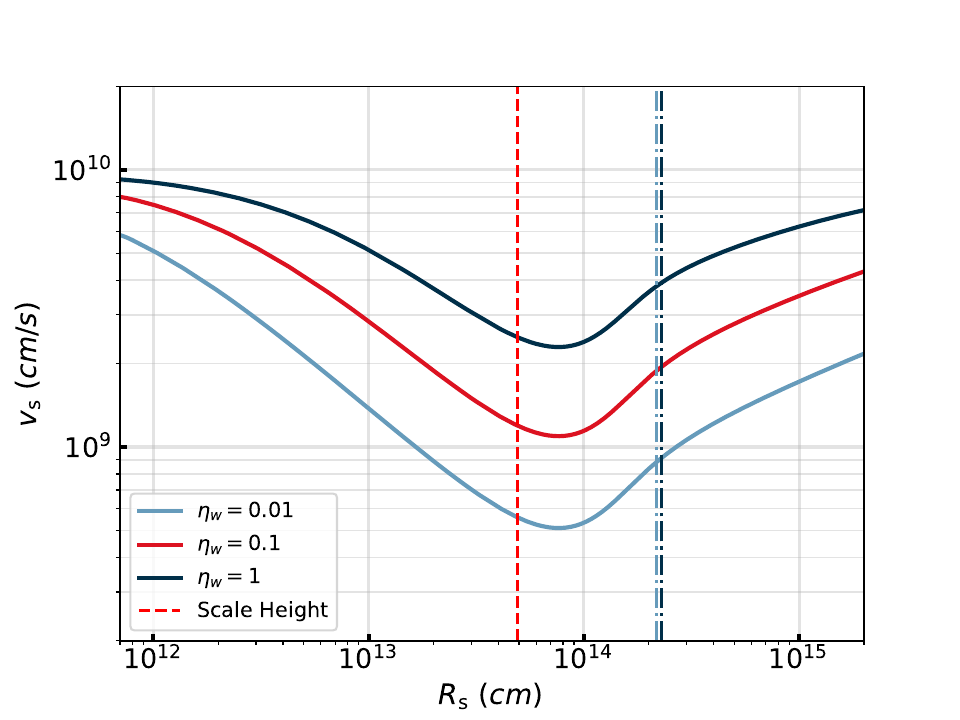}
\includegraphics[width=0.5\linewidth,height=0.4\linewidth]{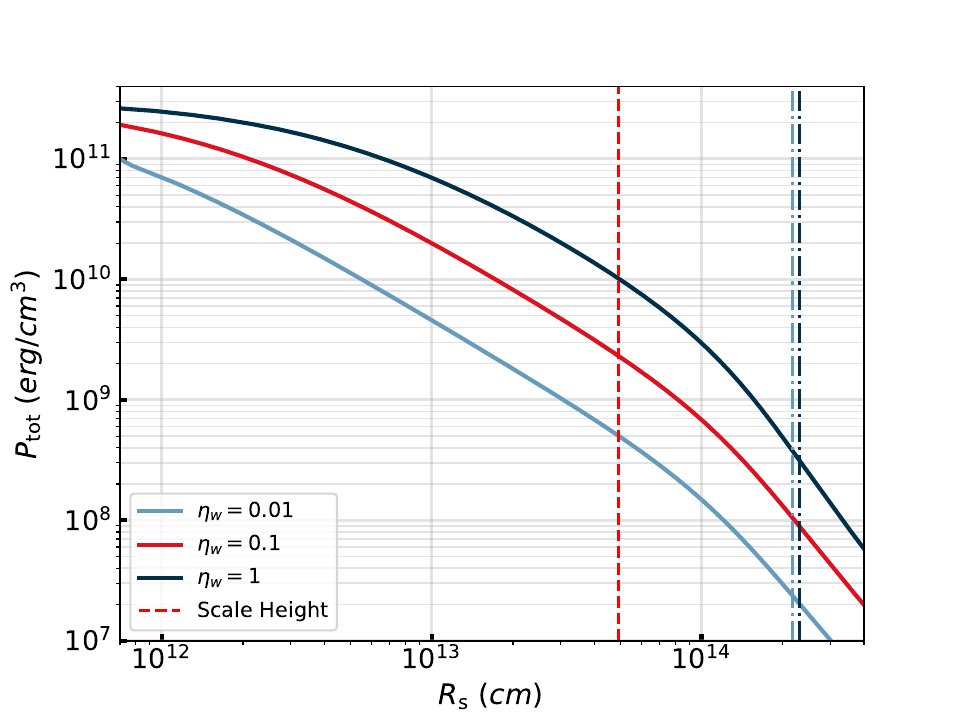}
\begin{center}
\includegraphics[width=0.5\linewidth,height=0.4\linewidth]{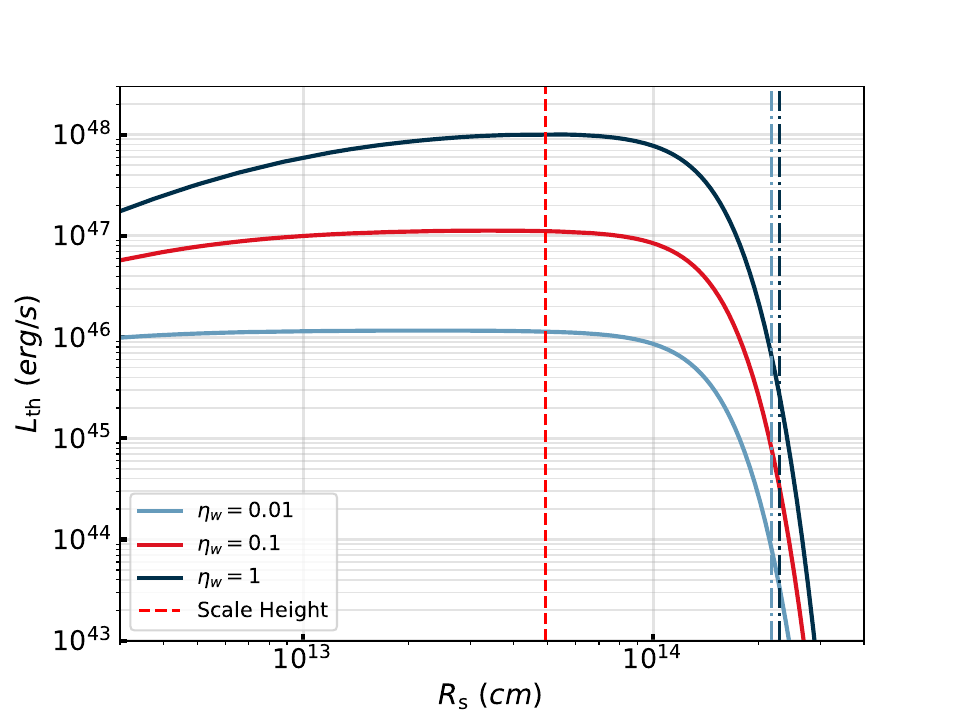}
\caption{The evolution of the forward shock velocity $v_{\rm s}$ (top-left), the total pressure of the shocked wind $P_{\rm tot}$ (top-right), and the thermal energy injection rate of shocked gas $L_{\rm th}$ (bottom), in the case of $M_{\rm SMBH}=10^7M_{\odot}, r=10^3R_{\rm sch}$. Solid lines in different colors represent different $\eta_{\rm w}$ selections. The vertical red dashed line represents the scale height of the AGN disk. The vertical dark-blue, blue, and red dot-dashed line represents the breakout radius $R_{\rm br}$ in each $\eta_{\rm w}$ selection. Note that the red dot-dashed line is overlapped by the dark-blue line and becomes invisible. }
\label{fig_dyn}
\end{center}
\end{figure*}
To calculate the pressure $P_{\rm tot}$, we have the equation for the total internal energy rate of Region (b),
\begin{equation}
\label{eq_eng}
     \frac{dE_{\rm tot}}{dt}=\frac{1}{2}\dot{M}_w(1-\frac{v_{\rm rs}}{v_{\rm w}})(v^2_{\rm w}-v'_{\rm sw})-4\pi R^2_{\rm s}P_{\rm tot}v_{\rm s}.
\end{equation}
The first term on the right-hand side of the above equation represents the injection rate of matter thermal energy in Region (b) (the difference between the wind kinetic power and shocked wind kinetic power). The second term is the energy loss due to the expansion. We also have the energy-pressure relation for the ideal gas,
\begin{equation}
\label{eq_pe}
    E_{\rm tot}=\frac{3}{2}P_{\rm tot}V_{\rm sw}=2\pi P_{\rm tot}(R^3_{\rm s}-R^3_{\rm rs}),
\end{equation}
with the volume of $V_{\rm sw}=4\pi /3(R^3_{\rm s}-R^3_{\rm rs})$ . Notice that Eq.(\ref{eq_eng}) and Eq.(\ref{eq_pe}) are the simplified equations, the original version is a set of differential equations describing the energy evolution of protons, electrons, and photons respectively. The original energy-pressure relation also includes that of photons. For simplification, We adopt this form in calculations, but the final result will not deviate so much. One can see the detailed argument in Appendix~\ref{sec_apx2}. When the evolution of $v_{\rm s}$ is obtained, we can also estimate the thermal energy injection rate of the forward shock by $L_{\rm th}=4\pi R^2_{\rm s}u_{\rm s}v_{\rm s}$, where $u_{\rm s}=\frac{3}{2}n_{\rm s}k_{\rm b}T_{\rm s}=\frac{9}{32}\rho_{\rm s}v^2_{\rm s}$ is the thermal energy density of the shocked disk gas, $T_{\rm s}=3m_{\rm p}v^2_{\rm s}/16k_{\rm b}$ is the temperature of the shocked disk gas, $\rho_{\rm s}=4\rho_{\rm disk}$ is the density of the shocked disk gas.

By equating the radiation-diffusive velocity and the shock velocity, one can obtain the characteristic optical depth $\tau_{\rm c}=c/v_{\rm s}$.
The shock will break out at the radius $R_{\rm br}$ where $\tau_{\rm c}=\tau_{\rm \rm m}$, $\tau_{\rm m}\simeq(\rho_{\rm disk}/m_{\rm p})\sigma_{\rm T}R$ is the material optical depth ahead of the shock~\citep{wang2019transient}. When $R_{\rm s} > R_{\rm br}$, the shock becomes collisionless and the particle acceleration starts to occur.

\subsection{Solutions for Dynamic Equations}

\begin{figure*}[t]
\includegraphics[width=0.5\linewidth,height=0.4\linewidth]{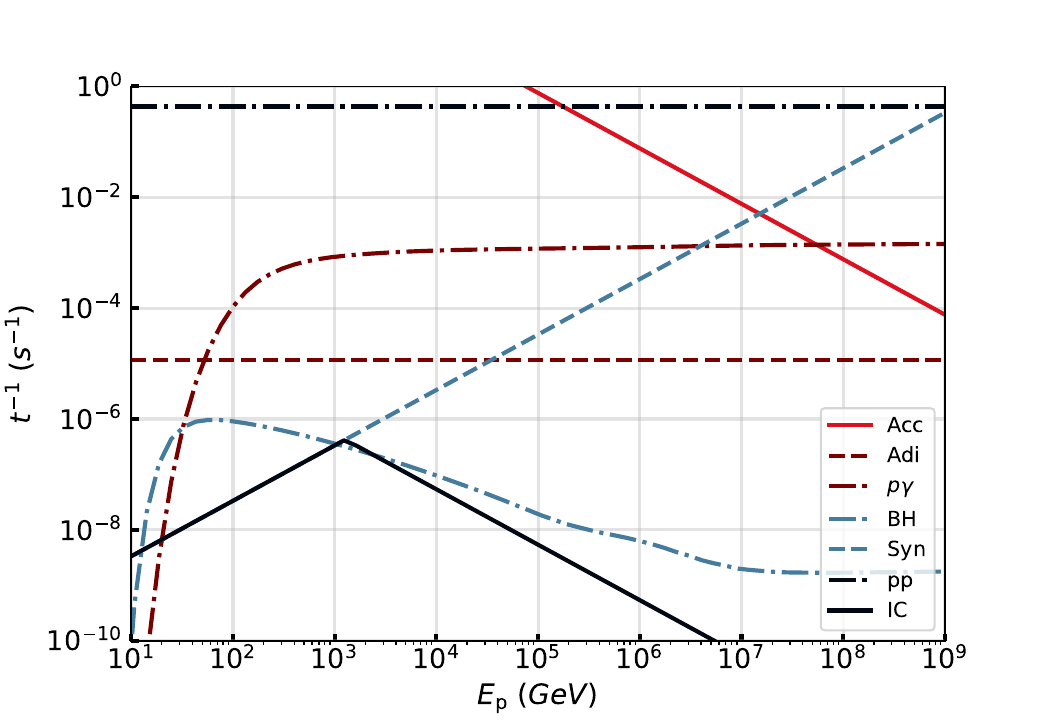}
\includegraphics[width=0.5\linewidth,height=0.4\linewidth]{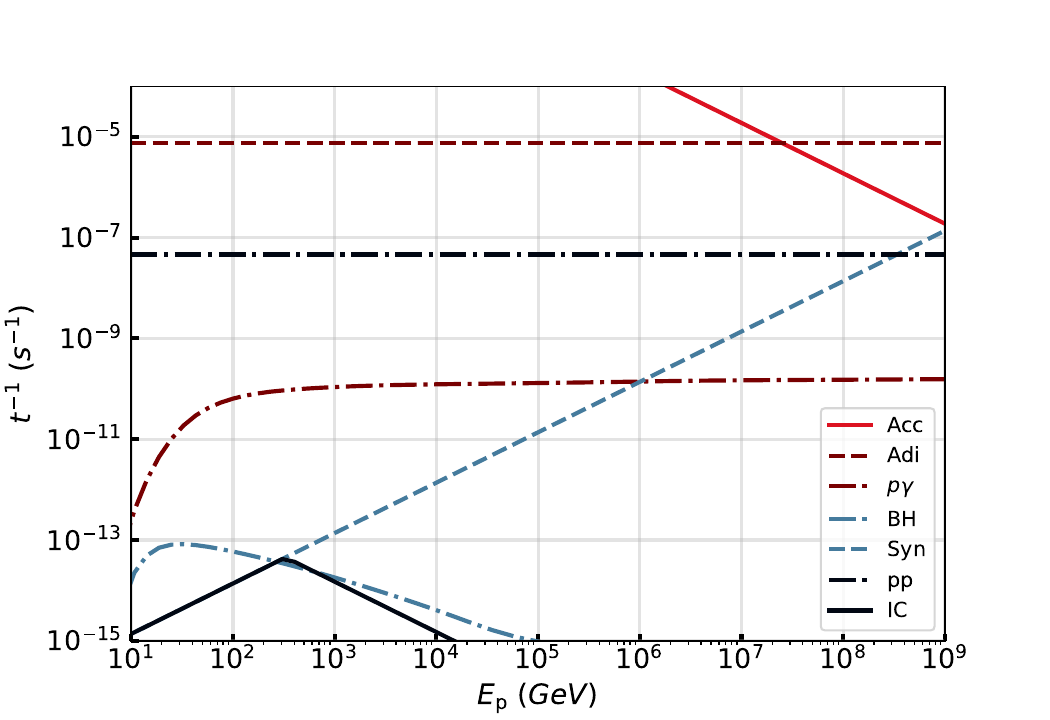}
\caption{The timescales for various processes of protons when the forward shock is at  $R_{\rm s}=1\times10^{14} \rm cm$ (left) and $R_{\rm s}=3\times10^{14} \rm cm$ (right) under the case of $M_{\rm SMBH}=10^7M_{\odot}, r=10^3R_{\rm sch}$, where $R_{\rm br}\sim2\times10^{14}\rm cm$. The maximum proton energy can be obtained by equating the timescale of the acceleration and the most effective cooling process. Note that the particle acceleration in the left panel (prior to shock breakout) is not allowed, and it is solely used for comparing the timescales of various processes at different radii.}
\label{fig_time}
\end{figure*}

We set the wind initially launched from the location $R=H/100$ to the merged BH in the midplane. Based on Eq.(\ref{eq_mov})-Eq.(\ref{eq_pe}), we can obtain the evolution of $v_{\rm s}$, $P_{\rm tot}$ and $L_{\rm th}$, we show our results in Fig~\ref{fig_dyn} with $M_{\rm SMBH}=10^7M_{\odot}, r=10^3R_{\rm sch}$, and diverse $\eta_{\rm w}$. We can see that the forward shock decelerates at the beginning and, when it passes through the scale height of the disk, it will continue to accelerate even beyond the breakout radius. The injected thermal luminosity is almost unchanged before the scale height, and then it becomes decreasing. When the shock reaches the breakout radius, $L_{\rm th}$ is relatively low compared to the initial value. We introduce the factor $\epsilon=0.1$ to represent the fraction of thermal luminosity converted to cosmic-ray luminosity~\citep{caprioli2014simulations}, that is, $L_{\rm cr}=\epsilon L_{\rm th}$.

In our solutions, the shock breakout radius is always larger than the scale height of the disk, i.e. $R_{\rm br} > H$. \cite{kimura2021outflow} proposed an analytic formula to estimate the breakout time: $t'_{br}\simeq\left(\frac{\rho_{\rm disk}H^5}{0.53L_{\rm w}}\right)^{1/3}$. In the case of $\eta_{\rm w}=0.1$, this formula gives $t'_{\rm br}\simeq 80 h$, which is comparable to our result $t_{\rm br}=44 h$.

\section{Neutrino Production}
\label{sec_np}
After the shock breakout, particles are accelerated at the forward shock front and produce high-energy neutrinos through hadronic processes. We consider that the accelerated protons follow the power-law distribution, namely, $dN_{\rm p}/dE_{\rm p} \propto E^{-s}_{\rm p}$, where $E_{\rm p}$ is the proton energy, $s=2$ is the proton spectrum index~\citep{blandford1987particle,malkov2001nonlinear}. The accelerated protons will interact with the shocked gas ($pp$ process) or the thermal photons produced by the shocked disk gas ($p\gamma$ process) and thus produce high-energy neutrino. Since the energy of secondary neutrinos comes from the parent protons, we can calculate the neutrino flux by~\citep{murase2016hidden}
\begin{equation}
\label{eq_lm}
    \epsilon^2_{\nu}F_{\nu}=\frac{3K}{4(1+K)}\frac{\zeta_{{\rm p}} L_{{\rm cr}}}{ 4\pi D^2_{{\rm L}}{\rm ln}(E_{\rm p,max}/E_{\rm p,min})},
\end{equation}
where $\epsilon_{\nu}=0.05E_{\rm p}$ is the neutrino energy, $D_{\rm L}$ is the the luminosity distance, $ {\rm ln}(E_{\rm p,max}/E_{\rm p,min})$ is the  normalized factor. $K=2$ for $t^{-1}_{\rm pp} > t^{-1}_{\rm p\gamma}$, $K=1$ for $t^{-1}_{\rm pp} < t^{-1}_{\rm p\gamma}$, where $t_{\rm pp}$ and $t_{\rm p\gamma}$ are timescales for $pp$ process and $p\gamma$ process. The time scale for $pp$ process is
\begin{equation}
t^{-1}_{\rm pp}=0.5\sigma_{\rm pp}n_{\rm s}c, 
\end{equation}
where $\sigma_{\rm pp}\sim5\times10^{-26} cm^{-2}$ is the cross section for $pp$ process, $n_{s}=4n_{\rm disk}=4\rho_{\rm disk}/m_{\rm p}$ is the number density of shocked disk gas~\citep{stecker1968effect,murase2007high}.The time scale for $p\gamma$ process is
\begin{equation}
t^{-1}_{\rm p\gamma}=\frac{c}{2\gamma^2_{\rm p}}\int_{\tilde{E}_{\rm th}}^{\infty}d\tilde{E}\sigma_{\rm p\gamma}(\tilde{E})\kappa_{\rm p\gamma}(\tilde{E})\tilde{E}\int_{\tilde{E}/2\gamma_{\rm p}}^{\infty}dE_{\gamma}E_{\gamma}^{-2}\frac{dN_{\gamma}}{dE_{\gamma}},
\end{equation}
where $\sigma_{\rm p\gamma}$ is the cross section for $p\gamma$ process~\citep{patrignani2016review}, $\tilde{E}$ is the energy of the photon in the rest frame of the proton, $\tilde{E}_{\rm th}\simeq 145 \rm MeV$ is the threshold energy, $\gamma_{\rm p}=E_{\rm p}/m_{\rm p}c^2$ is the Lorentz factor of the proton, and $dN_{\gamma}/dE_{\gamma}$ is the photon spectrum.

$\zeta_{\rm p}$ in Eq.(\ref{eq_lm}) represents the efficiency of $pp$ and $p\gamma$ reactions~\citep{murase2008prompt}, which can be estimated by $\zeta_{\rm p}=(t^{-1}_{\rm pp}+t^{-1}_{\rm p\gamma})/(t^{-1}_{\rm cool})$, where $t^{-1}_{\rm cool}=t^{-1}_{\rm pp}+t^{-1}_{\rm p\gamma}+t^{-1}_{\rm BH}+t^{-1}_{\rm syn}+t^{-1}_{\rm IC}+t^{-1}_{\rm adi}$ is the total cooling timescale for protons. The accelerated protons will lose energy via different channels besides $pp$ or $p\gamma$ processes, including adiabatic cooling with timescale $t_{\rm adi}=R_{\rm s}/v_{\rm s}$, synchrotron radiation cooling with timescale
\begin{equation}
    t_{\rm syn}=\frac{6\pi m^4_{\rm p}c^3}{\sigma_{\rm T}m_{\rm e}B^2E_{\rm p}},
\end{equation}
where the magnetic field strength can be estimated by $B=\sqrt{8\pi\epsilon_{\rm B}u_{\rm s}}=\sqrt{12\pi\epsilon_{\rm B}n_{\rm s}k_{\rm b}T_{\rm s}}$, $\epsilon_{\rm B}=0.1$.
The timescale of inverse-Compton scattering is~\citep{denton2018exploring}:

\begin{equation}
	t_{\rm IC}=\begin{cases}
		\frac{3m_{p}^{4}c^3}{4\sigma _Tm_{e}^{2}n_{\gamma}E _{\gamma}E _{\rm p}},&E _{\gamma}E _p<m_{p}^{2}c^4,\\
  \\
		\frac{3E _{\gamma}E _p}{4\sigma _Tm_{e}^{2}n_{\gamma}c^5}\,\,  ,&E _{\gamma}E _p>m_{p}^{2}c^4.\\
	\end{cases}
\end{equation}
where $n_{\gamma}$ is the number density of the thermal photon produced by shocked disk gas, it can be estimated by $n_{\gamma}=\epsilon_{\rm e}u_{\rm s}/E_{\gamma}$, where $E_{\gamma}=2.7k_{\rm b}T_{\rm s}$ is the typical thermal photon energy, $\epsilon_{e}=0.1$. Also, protons may take the Bethe–Heitler process with thermal photons, which timescale is~\citep{denton2018exploring},
\begin{equation}
    t_{\rm BH}=\frac{E_{\rm p}\sqrt{m^2_{p}c^4+2E_{\rm p}E_{\rm \gamma}}}{2n_{\gamma}\sigma_{\rm BH}m_{\rm e}c^3(E_{\rm p}+E_{\gamma})}.
\end{equation}
The cross-section of the Bethe-Heitler process can be written as
\begin{equation}
   \sigma_{\rm BH}=\alpha r^2_{\rm e}[\frac{28}{9}  {\rm ln}(\frac{2E_{\rm p}E_{\gamma}}{m_{\rm p}m_{\rm e}c^4})-\frac{106}{9}], 
\end{equation}
where $\alpha$ is the fine structure constant, $r_{\rm e}$ is the classical electron radius.

\begin{figure}
    \centering
    \includegraphics[width=1.0\linewidth,height=0.8\linewidth]{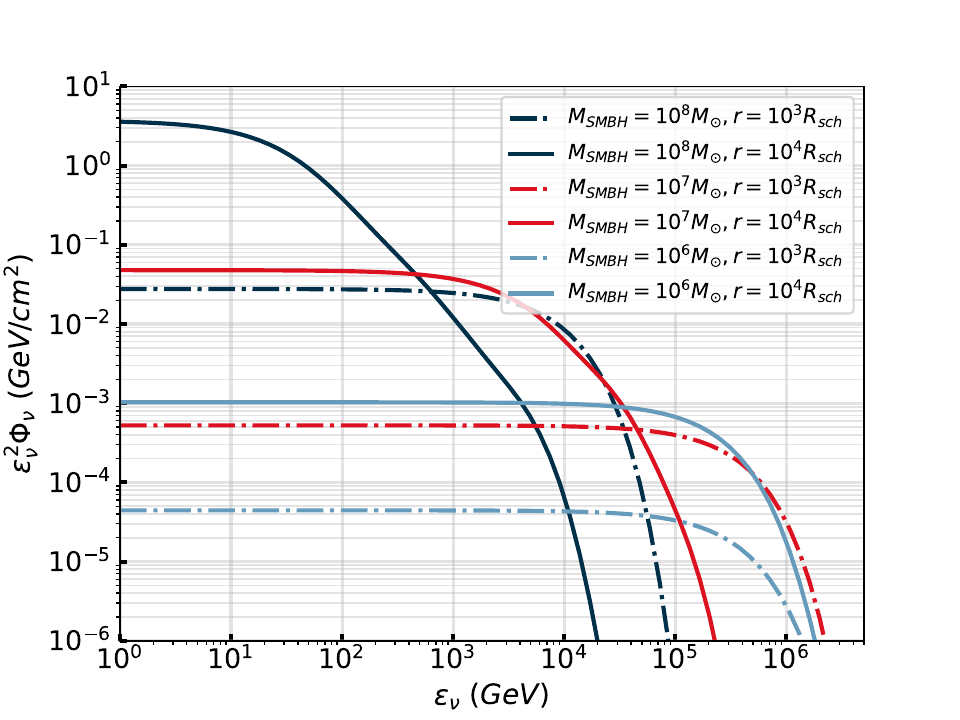}
    \caption{The all flavor neutrino fluence for an individual event occurring at $D_{\rm L}=10 ~\rm Mpc$ for the diverse SMBH masses ($M_{\rm SMBH}$) and locations of merged BH from the SMBH (r). $\eta_{\rm w}$ is fixed as 0.1.}
    \label{fig_flc}
\end{figure}

For the normalized factor in Eq.(\ref{eq_lm}), we set $E_{\rm p, min}=1 \,\rm TeV$ and the $E_{\rm p, max}$ can be obtained by equating the accelerated timescale $t_{acc}=20E_{\rm p}c/3eBv_{\rm s}$ with the minimum cooling timescale, where $e$ is the charge of the electron. In Fig.~\ref{fig_time}, we show the timescales for various processes when the forward shock expands at $R_{\rm s}=1\times10^{14} \rm cm$ and $R_{\rm s}=3\times10^{14} \rm cm$ (before and beyond $R_{\rm br}$ ) in the case of $M_{\rm SMBH}=10^7M_{\odot}, r=10^3R_{\rm sch}$. We notice that the $pp$ process is generally more efficient than the $p\gamma$ process.  For all considered cases, we summarize that the maximum proton energy after the shock breakout is in the range of 10 TeV to 20 PeV.

\section{Neutrino fluence, light curves, and Diffuse Neutrino Emission}
\label{sec_res}
\subsection{Individual Fluences, Light curves, and Expected Numbers}
The neutrino fluence for an individual event can be obtained by
\begin{equation}
    \epsilon^2_{\nu}\phi_{\nu}\simeq\int_{t_{\rm br}}^{\infty}\epsilon^2_{\nu}F_{\nu}dt,
\end{equation}
where $\epsilon^2_{\nu}F_{\nu}$ can be obtained through Eq.(\ref{eq_lm}). In Fig.~\ref{fig_flc}, we show the all-flavor neutrino fluence under various conditions, where the luminosity distance of the SMBH is $D_{\rm L} =10 ~\rm Mpc$ and $\eta_{\rm w}=0.1$. Based on the effective area $A_{\rm eff}(\epsilon_{\nu})$ of the IceCube detector, we also estimate the expected neutrino detection number~\citep{aartsen2020time},
\begin{equation}
    N_{\nu}=\int A_{\rm eff}(\epsilon_{\nu})\phi_{\nu} d\epsilon_{\nu},
\end{equation}
and present them in Table 1.

\begin{deluxetable}{cc}[t] 
\label{Tab:2}
\tablecaption{Number of detected $\nu_{\mu}$ from individual event at 10 Mpc by IceCube.}
\tablecolumns{2}
\tablewidth{0pt}
\tablehead{
\colhead{Models} &
\colhead{Detected Number}
}
\startdata
$10^8M_{\odot},10^3R_{sch}$ & $0.137$ \\
$10^8M_{\odot},10^4R_{sch}$ & $0.687$ \\
$10^7M_{\odot},10^3R_{sch}$ & $0.01$ \\
$10^7M_{\odot},10^4R_{sch}$ & $0.163$  \\
$10^6M_{\odot},10^3R_{sch}$ & $0.0007$ \\
$10^6M_{\odot},10^4R_{sch}$ & $0.017$ 
\enddata

\end{deluxetable}

\begin{figure}[t]
    \centering
    \includegraphics[width=1.0\linewidth,height=0.8\linewidth]{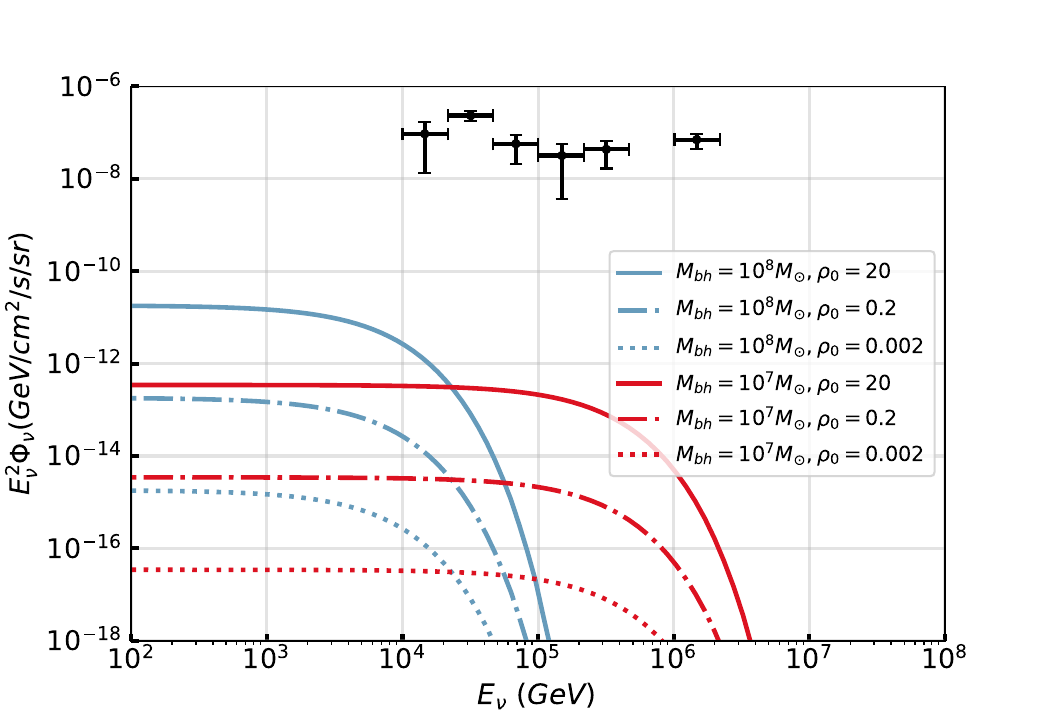}
    \caption{The contribution to the diffuse neutrino background by merged BH winds with different BBH merging rate densities, where $r=10^3 R_{\rm sch}$, $\eta_{\rm w}=0.1$. The black data points are observed astrophysical diffuse neutrino flux measured by the IceCube detector~\citep{aartsen2015search}. }
    \label{fig_diff}
\end{figure}

\begin{figure*}[t]
\begin{center}
\includegraphics[width=0.8\textwidth]{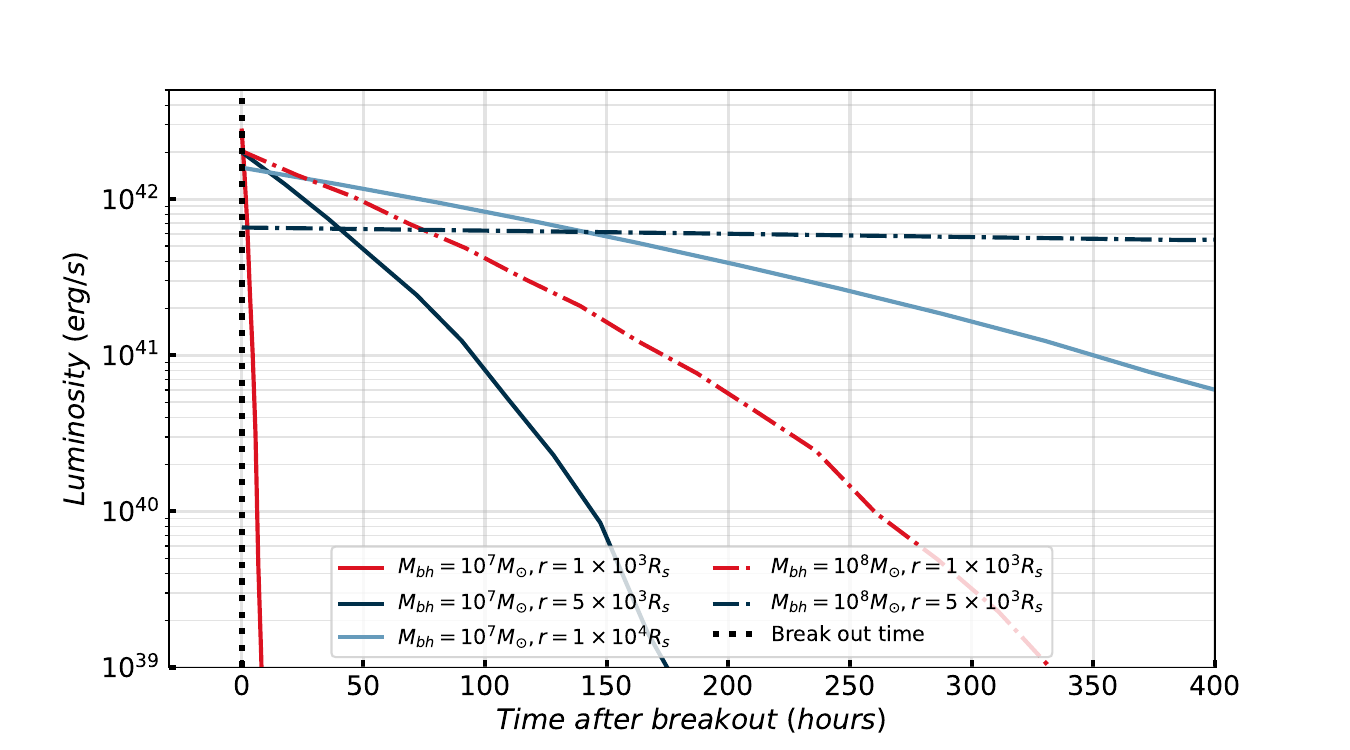}
\includegraphics[width=0.8\textwidth]{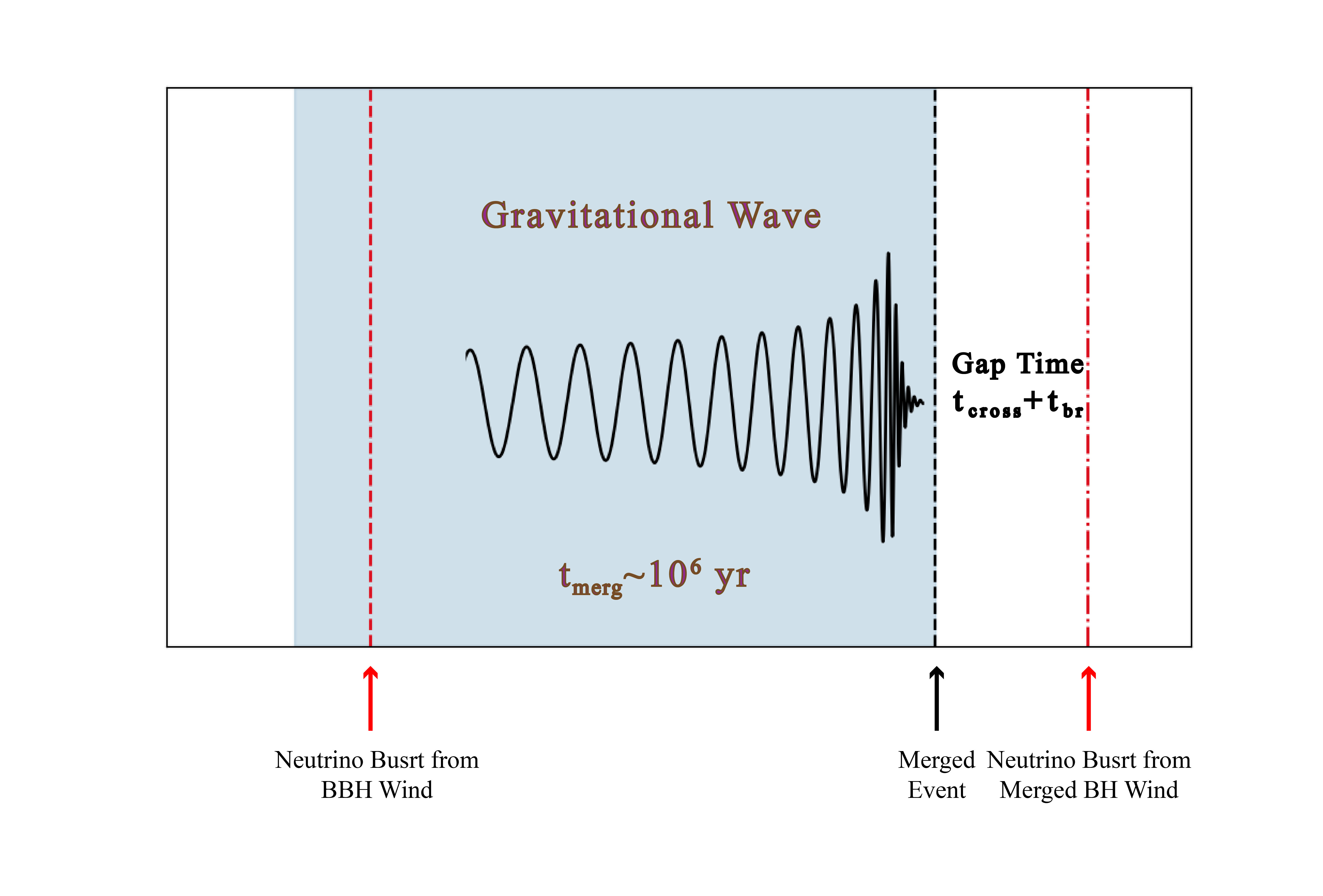}
\caption{Top panel: The light curves for neutrino bursts under various conditions. The vertical dotted line represents the shock breakout time (as the initial time). Bottom panel: Schematic light curve for the neutrino/EM and GW joint observation. There might be twice neutrino/EM bursts in the premerger phase and postmerger phase, respectively. The shaded region represents the gravitational wave duration time. }
\label{fig_lt}
\end{center}
\end{figure*}

From Fig.\ref{fig_flc}, we can see that neutrino fluence roughly increases with the mass of SMBH and radial location $r$. This is because the forward shock can exist longer after shock breakout when $M_{\rm SMBH}$ or $r$ are larger. The shock can carry more energy, and neutrino production can be more efficient, leading to higher neutrino fluence. This feature is also illustrated by the light curves shown in Fig.~\ref{fig_lt}. One can see from the light curves that the neutrino burst persists for several hours when $M_{\rm SMBH}$ or $r$ is small, while it can persist for even more than 400 hours when $M_{\rm SMBH}$ or $r$ is large. We can also find that the neutrino cutoff energy is located at the lower energy when the fluence is higher. In the case of $M_{\rm SMBH}=10^7M_{\odot}, r=10^3R_{\rm sch}$, the neutrino cutoff energy is located at several hundred TeV, while for $M_{\rm SMBH}=10^8M_{\odot}, r=10^4R_{\rm sch}$ it is at several ten GeV. The maximum proton energy will be suppressed for a more massive SMBH or a further location to SMBH because of more significant cooling processes for protons, inducing a smaller maximum proton energy and a smaller neutrino cutoff energy.

\subsection{ Diffuse Neutrino Emission}
The diffuse neutrino emission can be obtained through~\citep{razzaque2004tev,xiao2016high}
\begin{equation}
    E^2_{\nu}\Phi_{\nu}(E_{\nu})=\frac{c}{4\pi H_0}\int_0^{z_{{\rm max}}} 
    {\frac{\rho_{0}f(z)\epsilon^2_{\nu}\phi_{\nu}(\epsilon_{\nu})}{(1+z)\sqrt{\Omega_{{\rm m}}(1+z)^3+\Omega_{\Lambda}}}} dz,
\end{equation}
where $z$ is the redshift, $E_{\nu}=\epsilon_{\nu}/(1+z)$ is the observed neutrino energy, $H_0=67.8~ \rm km/s$ is the Hubble constant,
$\Omega_{\rm m}=0.308$, $\Omega_{\Lambda}=0.692$ are the cosmology constants~\citep{ade2016planck}. $f(z)$ is the redshift evolution factor and is roughly taken as the same as for short gamma-ray bursts because both are binary compact object mergers, that is, $f(z)=[(1+z)^{5.7\eta}+(\frac{1+z}{0.36})^{1.3\eta}+(\frac{1+z}{3.3})^{-9.5\eta}+(\frac{1+z}{3.3})^{-24.5\eta}]^{1/\eta}$ with $\eta=-2$~\citep{sun2015extragalactic}. The density of BBH merging rate within AGN disks $\rho_0$ is still quite uncertain, and we adopt $\rho_{0} \simeq [0.002, 20]~\rm Gpc^{-3}yr^{-1}$ proposed by~\cite{grobner2020binary}. The diffuse neutrino emissions produced by the merged BH wind are shown in Fig.~\ref{fig_diff}. We find that neutrino production from the remnant BH wind has little contribution to the diffuse neutrino background due to the low event rate density, but the contribution can be improved if the larger $\eta_{\rm w}$ and location $r$ from SMBH are involved. In addition, the neutrino from the remnant BH wind can still be identified from the near individual source as shown in Table~\ref{Tab:2}.

\subsection{Multimessenger Observations}
Although the diffuse neutrino contribution from merged BH winds is relatively low, the neutrino fluence for an individual nearby event is still considerable, even compared to that of other neutrino-produced activities on the disk, such as supernova explosions~\citep{zhu2021thermonuclear,Zhou2023a}, choked gammar-ray bursts~\citep{zhu2021high}, and jet powered by the merged BH~\citep{zhou2023b}.

Besides, there would be a detectable electromagnetic signal (thermal emission of the shocked gas) after the shock breakout, which luminosity can be estimated by $L_{\rm br}\simeq 7.1\times10^{45}~{\rm erg/s}(\frac{\rho_{\rm disk}}{\rm 10^{-9}g~cm^-{3}})(\frac{H}{10^{13.5}{\rm cm}})^2(\frac{v_{\rm s}}{10^9{\rm cm~s^{-1}}})^3 $~\citep{kimura2021outflow}. Such a signal may be distinguished from the background AGN radiation becoming an EM transient with the neutrino signal simultaneously. The BBH merger would produce detectable GW signals for Ligo/Virgo, thus we can expect a joint light curve for GW, EM, and neutrino, as illustrated in Fig.~\ref{fig_lt}. Notice that there might be twice neutrino/EM bursts in the premerger and postmerger phases, respectively. The former is produced by the BBH wind, which can be modeled by the same method as in Sections \ref{sec_dyn} and \ref{sec_np}. After the BBH merger, the kicked newborn remnant BH can spend a time of $t_{\rm cross}\simeq R_{\rm cav}/v_{\rm kick}$ to get out of the cavity formed by premerger BBH winds, where $R_{\rm cav}\sim R_{\rm Hill}$ is the radius of the cavity. Therefore, there will be a time delay between the GW signal from the BBH merger and the second EM/neutrino transient, i.e., $t_{\rm delay}=t_{\rm cross}+t_{\rm br}$. In the case of $M_{\rm SMBH}=10^7M_{\odot}, r=10^3R_{\rm sch}$, $ v_{\rm kick}=10^7 {\rm cm/s}$, the gap time is $\sim 60$ days.

\section{Summary}
\label{sec_sum}
The BBH within the AGN disks would merge and leave the remnant BH. The premerger outflows can create a cavity around the merged BH. When the newborn merged BH is kicked out of the cavity into a gas-abundant disk environment, it will accrete the disk gas in the way of the BHL accretion and power a fast wind. Such fast outflow will drive a forward shock in the disk gas and thus accelerate the protons and electrons. The accelerated protons will subsequently interact with photons/matters of the shocked gas to produce high-energy neutrinos. In this paper, we investigate the detailed dynamic evolution of the interaction between the remnant BH wind and the disk medium, calculate the neutrino production and expected detection numbers for an individual event, and estimate the contribution of these events to the diffuse neutrino background.

The stellar-mass BBH merger events have become the most important detection targets for ground-based GW detectors including Ligo, Virgo and KAGRA~\citep{PhysRevX.9.031040}. In addition to GW bursts, BBH mergers embedded in AGN disks are expected to produce potential EM and neutrino signals, and thus the multimessenger study for BBH mergers in AGN disks is very desirable. Especially, high-energy neutrinos will be an ideal and unique messenger for studying BBH mergers within AGN disks, as they can easily escape from the complicated environments of AGNs. Recently, \cite{Abbasi_2023} performed the high-energy neutrino searches for GW events and no significant neutrino emission was found, which is consistent with our results for their far luminosity distances and low expectations of neutrino fluxes. The neutrino follow-up search on the candidate optical counterpart of the BBH merger event GW190521 was carried out as well. However, the relatively far distance of GW190521, i.e., $3931\pm 953\,\rm Mpc$~\citep{graham2020candidate}, makes the expected neutrino fluence low and can not provide the useful constraints currently. Future GW detectors will further significantly increase the detection number of BBH mergers, which is expected to reach $10^5-10^6\,\rm yr^{-1}$~\citep{2020JCAP...03..050M}, making them one of the most important astrophysical sources. The association between GWs and high-energy neutrinos can be expected.

\section*{Acknowledgments}
We acknowledge support from the National Natural Science Foundation of China under grant No.12003007 and the Fundamental Research Funds for the Central Universities (No. 2020kfyXJJS039).

\appendix

\section{The modified $\alpha$-disk model}
\label{sec_apx1}

To describe the midplane of the disk, a modified $\alpha$-disk model is adopted~\citep{cantiello2021stellar} and we assume that all processes occur in the gas-pressure-dominated region. By considering the gravitational stability, the disk is separated into two regions by a characteristic radius,
\begin{equation}
R_Q\simeq0.02\left(\frac{\kappa^3\alpha^7}{\mu^{12}}\right)^{2/27}\frac{M^{1/27}_8}{f^{8/27}_m}\ {\rm pc},
\end{equation}
where $\kappa\sim1$ is the opacity, $\alpha\sim0.1$ is the viscosity factor and $\mu\sim0.5$ is the molecular weight. $M_8=M_{\rm SMBH}/10^8M_{\odot}$ with the mass of central SMBH $M_{\rm SMBH}$ and the mass of the Sun $M_{\odot}$. At a distance $r \le R_{Q}$ from the SMBH, the Toomre parameter Q satisfies $Q\gtrsim 1$, suggesting that the disk is gravitationally stable. In such region, the midplane gas temperature and density can be described by
\begin{equation}
\label{eq_T1}
    T_{\rm mid}=3.7\times10^3\left(\frac{\kappa\mu}{\alpha}\right)^{1/5}\frac{f^{2/5}_mM^{7/10}_8}{r^{9/10}_{\rm pc}}\ {\rm K},
\end{equation}
\begin{equation}
\label{eq_r1}
\rho_{\rm mid}=1.5\times10^{-13}\frac{\mu^{6/5}f^{2/5}_mM^{19/20}_8} {\kappa^{3/10}\alpha^{7/10}r^{33/20}_{\rm pc}}\ {\rm g/cm^3},
\end{equation}
respectively, where $r_{\rm pc}=r/1 \ {\rm pc}$, $f_m\sim1$ is the dispersion factor. We also have the scale height for the disk,
\begin{equation}
\label{eq_h1}
    H\simeq\sqrt{\frac{2R_gT_{\rm mid}}{\mu\Omega}},
\end{equation}
where $R_g$ is the special gas constant defined as $R_g=k_b/m_{\rm p}$, $\Omega=\sqrt{GM_{\rm SMBH}/r^3}$ is the angular frequency at a distance r from the SMBH and surface density is
\begin{equation}
   \Sigma=2\rho_{\rm mid}H,
\end{equation}

At the region $r > R_{Q}$, the conventional $\alpha$-disk model should be modified by considering the gravity normal to the disk plane, momentum transfer, and possible thermal and momentum feedback from newly born/captured stars within the disk. The gravitational stability will be maintained with $Q\sim1$ by assuming a self-regulated star formation rate, which leads to the modified formulas at $r > R_{Q}$,
\begin{equation}
\label{eq_T2}
    T'_{\rm mid}=1.3\times10^ 4M^{2/3}_8\left(\frac{f_{\rm m}Q}{\alpha}\right)^{2/3}\ {\rm K},
\end{equation}
\begin{equation}
\label{eq_r2}
    \rho'_{\rm mid}=8.3\times10^{-16}\frac{M_8}{r^3_{\rm pc}Q} \ {\rm g/cm^3},
\end{equation}
Also, the scale height and surface density are
\begin{equation}
\label{eq_h2}
    H'=\frac{0.025r^{1/2}_{\rm pc}f^{1/3}_{\rm m}Q^{1/3}r}{\alpha^{1/3}M^{1/6}_8}\ {\rm cm},
\end{equation}
\begin{equation}
\label{eq_s2}
    \Sigma'=\frac{180M^{5/6}_8f^{1/3}_{\rm m}}{r^{2/3}_{\rm pc}Q^{2/3}\alpha^{1/3}}\ {\rm g/cm^3},
\end{equation}

When the location of the merged black hole is in the region of $r \le R_{Q}$, we use the formulas Eq.(\ref{eq_T1})--Eq.(\ref{eq_h1}) to describe the midplane properties of the disk; otherwise, the formulas Eq.(\ref{eq_T2})--Eq.(\ref{eq_s2}) are adopted.

\section{simplication for dynamic equations}
\label{sec_apx2}
In this section, we give the original dynamical equations for the shocked wind region (Region (b)), our model is similar to that of ~\cite{liu2018can},\cite{wang2015probing} and \cite{wang2016cumulative}. The electrons, protons, and photons in Region (b) will interact with each other through various channels. First, protons and electrons will lose energy through free–free emission and synchrotron emission. In addition, protons and electrons may collide with each other to exchange energy, namely the Coulomb collision. Finally, they will change energy with photons through Compton scattering. By considering these particle interactions, the internal energy rates for protons, electrons, and photons are described by
\begin{equation}
\label{eq_p}
    \frac{dE_{\rm p}}{dt}=\frac{1}{2}\dot{M}_{\rm w}(1-\frac{v_{\rm rs}}{v_{\rm w}})(v^2_{\rm w}-v'_{\rm sw})-4\pi R^2_{\rm s}P_{\rm psw}v_{\rm s}-L_{\rm cou}-L_{\rm rad,p}-L_{\rm com,p}
\end{equation}
\begin{equation}
\label{eq_e}
      \frac{dE_{\rm e}}{dt}=\frac{1}{2}\dot{M}_{\rm w}(\frac{m_{\rm e}}{m_{\rm p}})(1-\frac{v_{\rm rs}}{v_{\rm w}})(v^2_{\rm w}-v'_{\rm sw})-4\pi R^2_{\rm s}P_{\rm esw}v_{\rm s}+L_{\rm cou}-L_{\rm rad,e}-L_{\rm com,e}
\end{equation}
\begin{equation}
\label{eq_ph}
         \frac{dE_{\rm ph}}{dt}=L_{\rm rad,p}+L_{\rm rad,e}+L_{\rm com,p}+L_{\rm com,e}-4\pi R^2_{\rm s}P_{\rm ph}v_{\rm s},
\end{equation}
where the first terms on the right hand of Eq.(\ref{eq_p}) and Eq.(\ref{eq_e}) are the injected thermal energy rate for protons and electrons. $L_{\mathrm{rad},i}$ and $L_{\mathrm{com},i}$ ($i=e,p$) are the radiation cooling rate and Compton cooling/heating rate for electrons and protons~\citep{sazonov2001gas,faucher2012physics}, while they are energy injection term for photons. $L_{\rm cou}$ is the energy changing rate through Coulomb collision. $4\pi R^2_{\rm s}P_{\rm j}v_{\rm s}$ (j=psw,esw,ph) is the expanding cooling term for protons, electrons, and photons. We also have the energy-pressure relations,
\begin{equation}
\label{ep1}
    E_{\rm p/e}=\frac{3}{2}P_{\rm p/e}V_{\rm sw}
\end{equation}
for gas, which comes from the ideal gas relation $P_{g}=nk_{\rm b}T=\frac{2}{3}u_{\rm g}$ with the energy density of the gas $u_{\rm g}$. Also, we have
\begin{equation}
\label{ep2}
    E_{\rm ph}=3P_{\rm ph}V_{\rm sw}
\end{equation}
for photons, which comes from the relation $P_{\rm ph}=\frac{1}{3}u_{\rm ph}$ for uniform photon fields.

In principle, one can solve the above equations together with Eq.(\ref{eq_mov}) to obtain the thermal energy/pressure evolution of three species of particles. However, we only care about the total pressure $P_{\rm tot}=P_{\rm psw}+P_{\rm esw}+P_{\rm ph}$. Therefore, we can simplify the equations Eq.(\ref{eq_p})--Eq.(\ref{eq_ph}) by adding them together. Then all the energy interaction terms will vanish and the injected power of protons will be much larger than that of electrons ($m_{\rm e}/m_{\rm p}\simeq1/1000$), which gives the following
\begin{equation}
     \frac{dE_{\rm tot}}{dt}=\frac{1}{2}\dot{M}_w(1-\frac{v_{\rm rs}}{v_{\rm w}})(v^2_{\rm w}-v'_{\rm sw})-4\pi R^2_{\rm s}P_{\rm tot}v_{\rm s}.
\end{equation}
It is just Eq.(\ref{eq_eng}). For the energy-pressure relationship, we notice that the right hand of Eq.(\ref{ep1}) and Eq.(\ref{ep2}) is nearly comparable with only twice the difference. Namely, the pressure provided by gas/photon is comparable when they have the same energy. Therefore, we uniformly use Eq.(\ref{ep1}) for simplification.


\bibliography{reference}{}
\bibliographystyle{aasjournal}



\end{document}